\documentclass[aps,prl,reprint,superscriptaddress,nofootinbib]{revtex4-2}

\usepackage{amsmath}
\usepackage{graphicx}
\usepackage{booktabs}
\usepackage{siunitx}
\DeclareSIUnit\year{yr}
\newcolumntype{d}{S[table-format=2.2e1]}
\usepackage{chemformula}
\usepackage{url}
\usepackage{hyperref}

\begin{document}

\title{AI Hastens Limits to Exponential Growth}

\author{Robert T. Nachtrieb}
\affiliation{MIT Sloan School of Business}

\author{Steven J. Smith}
\affiliation{Blue Origin Space}

\date{2026-04-20}

\begin{abstract}
At any sustained positive growth rate of energy demand, depletion of all
terrestrial energy resources, including non-renewable deuterium fusion and
renewable solar, occurs within a remarkably compressed period. The time to
depletion is inversely proportional to the demand growth rate. Artificial
Intelligence (AI) has the potential to increase the growth rate of electricity
from three percent per year to fifteen percent per year, effectively collapsing
multi-millennial expansion timelines into decades. To grow after terrestrial
depletion will require capturing more of the sun's output than the earth's
cross-sectional area, eventually capturing the entire sun's output (Kardashev
Type~II civilization). Expansion beyond that threshold requires colonizing other
star systems. Simple algebraic models yield the main conclusions of the paper,
supported by a system dynamics simulation. This analysis reveals that even
unthinkably vast resources, such as total oceanic deuterium or the full
luminosity of the Sun, are decidedly finite when viewed through a logarithmic
lens. Uncertainties in the exact remaining resources of coal, oil, natural gas,
and uranium do not affect the conclusions of this paper, as the fundamental
physical limit is dictated by the geometry of expansion and the universal speed
of light.
\end{abstract}


\maketitle

%%=============================================================================
\section{Introduction}
%%=============================================================================

\subsection{Historical and Projected Demand Growth}

In 1964, Soviet astronomer Nikolai Kardashev proposed a seminal framework to
classify technologically advanced civilizations based on their mastery over
energy. This scale envisions three levels: Type~I (Planetary), harnessing all
available energy resources on its home planet; Type~II (Stellar), controlling
the total energy output of its host star, potentially via megastructures like a
Dyson sphere; and Type~III (Galactic), accessing and controlling energy on the
scale of its entire host galaxy~\cite{NK64}.

Kardashev's calculations demonstrated that even with a modest, sustained energy
demand growth rate of $\SI{0.01}{\per\year}$, a civilization would reach the
Type~II level in a remarkably short time of \SI{3200}{\year}~\cite{NK64}. This model
relies on the exponential growth formula:
\begin{equation}
  D(t) \approx D(0)\exp(r\,t)
\end{equation}
where $r$ [\si{\per\year}] is the constant growth rate, $t$ [\si{\year}] is time, and $D$ [\si{\exa\joule\per\year}] is the energy demand. However,
historical data indicates global energy consumption between 1964 and 2024 grew
at an average rate of $\SI{0.0214}{\per\year}$~\cite{EIA25}, with electricity
consumption more recently growing at $\SI{0.0289}{\per\year}$~\cite{Em26}, well
beyond Kardashev's estimates.

The recent emergence and rapid adoption of Artificial Intelligence (AI) has
further altered this calculus, representing a distinct phase shift in global
energy dynamics. Unlike historical industrial growth, which scaled with
population and physical labor, the energy requirements for AI have begun to
decouple from general economic trends. The exponential growth of AI is primarily
driven by continuous advancements in computing power, the expansion of scalable
cloud infrastructure, and the widespread availability of massive datasets used
to train increasingly complex algorithms. Businesses are heavily investing in
these technologies to gain a competitive edge through task automation, enhanced
predictive analytics, and highly personalized customer
experiences~\cite{Sab}. However, this rapid expansion requires staggering
amounts of energy because training large language models and processing
continuous, real-time inferences for billions of daily queries demand immense,
high-performance computational capacity~\cite{Cla26}. Consequently, the data
centers providing this necessary processing power are becoming significantly
larger and more power-intensive, drawing tens of \si{\giga\watt} of electricity to
sustain the advanced hardware and intensive cooling systems required for modern
AI workloads~\cite{IEA,Cla26}.

\begin{figure}[htbp]
  \centering
  \includegraphics[width=\columnwidth]{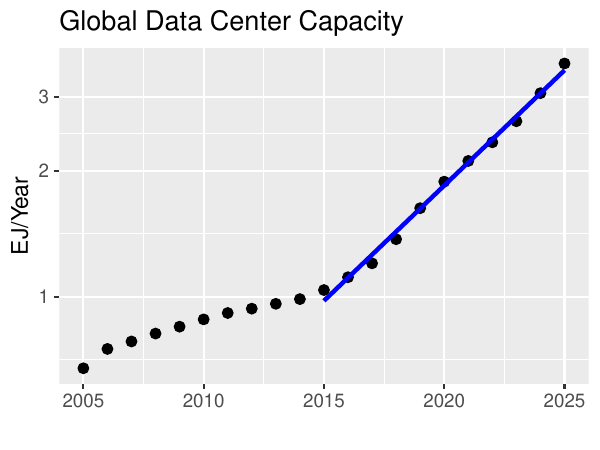}
  \caption{Global Computer Data Center Capacity~\cite{IEA}.}
  \label{fig:AI}
\end{figure}

\begin{figure}[htbp]
  \centering
  \includegraphics[width=\columnwidth]{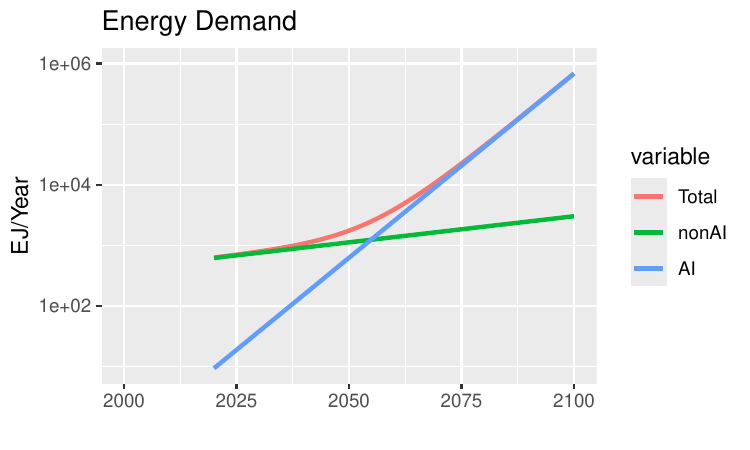}
  \caption{Near-term Demand Projection for non-AI and AI.}
  \label{fig:demand}
\end{figure}

While AI electricity consumption was only 1.5\% of the global total in 2024,
its power demand has grown at a rate of $\SI{0.127}{\per\year}$ since 2015,
accelerating to $\SI{0.15}{\per\year}$ over the last five years~\cite{IEA}
(Figure~\ref{fig:AI}). Projecting from this 2024 baseline, AI's electricity
demand is on track to achieve parity with the rest of the world's combined
consumption by approximately 2050 (Figure~\ref{fig:demand}). At this inflection
point, the global energy frontier will be dominated by the AI growth rate rather
than traditional industrial growth. This profound acceleration is the primary
driver shortening the timescales for planetary resource depletion explored in
the subsequent sections of this paper.

\subsection{Scope}

We model the balance between growing demand for energy and supply from all
available sources. As long as supply is available, growth in demand can
continue. This analysis begins within the energetic constraints of a
Kardashev Type~I (Planetary) civilization and models the transition into
Type~II (Stellar) and Type~III (Galactic) scales as demand inevitably exceeds
terrestrial limits.

We consider a system dynamics model of the energy portion of the industrial
system, including both non-renewable and renewable resources. We exclude from
scope several important topics, including population, non-energy portions of
the industrial system, other non-renewable resources, agriculture, pollution,
carbon emissions and climate change. While these factors are critical to
planetary health, they are treated here as separate from the fundamental
physical limit of energy availability and waste-heat dissipation.

We treat the rates of demand growth of AI and non-AI as exogenous inputs to
the system. Turliuk and Sterman~\cite{TS26} identify several reinforcing
feedback loops that will drive the growth rate of demand for AI. They include
effects of AI use on per capita gross world product, AI energy intensity, and
AI cost. Dynamics include efficiency rebound effects and a number of benefits
that accrue with short and long delays.

%%=============================================================================
\section{Resources}
%%=============================================================================

AI is powered by electricity generated here on Earth from both non-renewable
and renewable sources.

\subsection{Non-Renewable}

When we typically think of non-renewable energy resources on Earth, we think of
burning hydrocarbons, namely coal, oil, and natural gas. Today's nuclear power
plants use uranium-based fission to generate energy. However, for the purpose
of this analysis, the primary non-renewable resource of the advanced future is
nuclear fusion.

Nuclear fusion offers the promise of a seemingly limitless and clean energy
future by replicating the physical processes powering the sun and
stars~\cite{Prz23}. At the heart of this transformative technology is
deuterium, a naturally occurring ``heavy'' form of hydrogen serving as a
primary fuel component~\cite{ITER}.

Because deuterium is found abundantly in ordinary seawater, it provides what
appears to be an inexhaustible fuel supply for near-term human use. By
extracting this isotope from the oceans and fusing it under extreme conditions,
scientists aim to unlock a massive, sustainable source of power while producing
no greenhouse gases~\cite{Prz23,Smo}. The energy density of fusion is so high,
millions of times greater than chemical combustion, that it fundamentally shifts
the scale of available terrestrial energy.

However, from a planetary perspective, even fusion must be classified as a
non-renewable resource. This is because the total inventory of deuterium in
Earth's oceans, while vast, is a finite stock. Once the growth of demand begins
to scale exponentially, even this ``limitless'' resource faces a measurable
depletion horizon.

Table~\ref{tab:nonrenew} lists the major non-renewable energy resources
available. The essential observation is that deuterium for nuclear fusion far
exceeds any other non-renewable energy resource. While uncertainties may exist
about the exact quantity of remaining oil, for example, it doesn't matter: the
oil will be depleted long before we run out of deuterium.
Appendix~\ref{sect:deuterium} contains details of the deuterium resource
estimate.

\begin{table*}[htbp]
  \centering
  \sisetup{
    retain-unity-mantissa = true,
    table-align-exponent  = true,
  }
  \begin{tabular}{lS[table-format = 3.0]c@{\quad}ldc@{\quad}dc@{\quad}dcc}
    & {Current}  & & &             & & {Current}   & &             & \\
    & {Prod.}    & & & {Resources} & & {Cons.}     & & {Resources} & \\
    & {(\si{\exa\joule\per\year})} & & \si{unit} & {(\si{unit})} & & {(\si{unit\per\year})} & & {(\si{\exa\joule})} & \\
    \hline
    Coal        & 178  & [a] & \si{\mega.ton}       & 20.8e6  & [b] & 6.40e3 & [b] & 578e3   & & \\
    Oil         & 193  & [a] & \si{\mega.bbl}       & 6.16e6  & [b] & 35.6e3 & [b] & 33.4e3  & & \\
    Natural Gas &  48  & [a] & \si{\giga.\meter^3}  & 792e3   & [b] & 4.28e3 & [b] & 27.4e3  & & \\
    Uranium     &  31  & [a] & \si{ton}             & 10.7e6  & [c] & 59.0e3 & [c] & 5.62e3  & & \\
    Deuterium   &      &     & \si{\giga\mol}       & 22.7e9  & [d] &        &     & 15.8e12 & & [e] \\
    \hline
    Total       & 633  &     &                      &         &     &        &     & 15.8e12 & & \\
  \end{tabular}
  \par\vspace{4pt}
  \raggedright\footnotesize
  [a]~IEA World Energy Outlook 2025, p.\,414\quad
  [b]~IEA World Energy Outlook 2025, pp.\,433,435,436\quad
  [c]~IAEA Red Book 2025, p.\,112\quad
  [d]~\SI{33}{\gram\per\cubic\meter} in seawater; see Appendix~\ref{sect:deuterium}\quad
  [e]~\ch{6 _1^2D -> 2 _2^4He + 2 _0^1n + 2 _1^1p + 43.2 MeV}; see Appendix~\ref{sect:deuterium}
  \caption{Non-renewable energy resources are dominated by deuterium for nuclear
    fusion, more than twenty million times the combined resources of coal, oil,
    natural gas and uranium.}
  \label{tab:nonrenew}
\end{table*}

\subsection{Renewable}

We consider all renewable energy resources to trace back to the sun. Whether
the immediate mechanism of generation is photovoltaic (direct) or wind
(atmospheric pressure differentials), these sources are all derivatives of the
primary solar flux.

If humanity achieves a theoretical 100\% efficiency in capturing and converting
the sun's energy, neglecting for a moment the physical constraints of the
Shockley--Queisser limit for semiconductors~\cite{Sho61}, then the absolute
maximum available power at Earth's orbital radius is the solar flux multiplied
by the Earth's cross-sectional area~\cite{IPCC}.

This ``Solar Ceiling'' represents a hard physical limit for a strictly
terrestrial civilization. Once energy demand grows to meet this flux, no amount
of further terrestrial technological innovation can increase the supply of
renewable energy.

\begin{table}[htbp]
  \centering
  \caption{Solar parameters.}
  \label{tab:solar}
  \begin{tabular}{lllS[table-format=3.2e2,table-align-exponent=true]}
    \toprule
    Description          & Symbol & Units                    & {Value}            \\
    \midrule
    Sun luminosity       & $L$    & \si{\exa\joule\per\year} & 12e15              \\
    Solar power at Earth & $S$    & \si{\exa\joule\per\year} & 5.44e6             \\
    Earth--Sun distance  & $X_o$  & \si{\kilo\meter}         & 150e6              \\
    Earth radius         & $X_e$  & \si{\kilo\meter}         & 6.37e3             \\
    Geometric flux fraction & $\alpha$ & Dmnl               & {}                 \\
    \bottomrule
  \end{tabular}
\end{table}

\begin{align}
  \alpha &= \frac{\pi\,X_e^2}{4\pi\,X_o^2} \\
  S      &= L\,\alpha
\end{align}

For calculations using values above gives $\alpha \approx 4.53\times10^{-10}$.

%%=============================================================================
\section{Resource Depletion}
%%=============================================================================

To establish a mathematical foundation for our analysis, we assume AI-driven
electricity demand is the primary long-term driver of global energy consumption.
To build intuition regarding the velocity of this transition, we derive simple
estimates of resource depletion and physical thresholds under conditions of
constant exponential growth.

In all scenarios explored below, the time to depletion ($t$) is defined by a
fundamental inverse-growth relationship: $t = k/r$. Here, the depletion time
is directly proportional to a specific rate coefficient ($k$) and inversely
proportional to the demand growth rate ($r$).

This relationship illustrates the mathematical sensitivity of planetary
timelines to technological acceleration. While the rate coefficients ($k$)
remain fixed by physical laws and planetary geometry, the growth rate ($r$) is
a variable influenced by the pace of AI adoption. Consequently, as $r$ shifts
from historical averages to AI-accelerated levels, the timeframe for
transitioning from terrestrial to stellar-scale energy management is
significantly compressed.

\subsection{Demand}

The mathematical framework for this analysis relies on three primary equations
that define the evolution of global energy requirements. These metrics quantify
the transition from current consumption levels to planetary-scale resource
strain.

\begin{enumerate}

\item Instantaneous Demand $D(t)$:
\begin{equation}
  D(t) = D(0)\exp(r\,t)
\end{equation}
This expression models the temporal trajectory of energy needs. It establishes
that demand functions as a compounding variable where the rate of increase is
proportional to the current magnitude, resulting in a non-linear acceleration
of power requirements over time.

\item Cumulative Consumption $Q(t)$:
\begin{equation}
  Q(t) = \frac{D(0)}{r}\bigl[\exp(r\,t) - 1\bigr]
\end{equation}
This integral defines the total volume of energy extracted from the environment
over a given period. In a systems dynamics context, $Q(t)$ represents the total
drain on finite resource stocks; it illustrates how the area under an
exponential curve can exhaust even vast reserves within a limited number of
doubling periods.

\item Expansion Velocity $dD/dt$:
\begin{equation}
  \frac{dD}{dt} = r\,D(0)\exp(r\,t)
\end{equation}
The first derivative of demand with respect to time quantifies the speed of
expansion. This metric identifies the rate at which new energy capacity must be
brought online to satisfy growth, highlighting that the challenge is not only
the absolute magnitude of demand but the exponential acceleration of the
expansion rate itself.

\end{enumerate}

By establishing these mathematical relationships between growth rates and
resource consumption, we can project the chronological milestones of planetary
and stellar expansion. The following five cases represent a sequence of
escalating physical thresholds. These limits range from the exhaustion of
terrestrial fuel stocks to the fundamental thermodynamic and relativistic
constraints of the galaxy. Each occurs as the compounding energy demand
intersects a fixed physical constant.

\subsection{Case 1: Non-Renewable Supply Only}

This scenario models a system where demand is satisfied exclusively by finite
terrestrial energy stocks. Resource depletion occurs at the specific moment
cumulative energy extraction reaches the total initial resource stock. This
equality represents the physical exhaustion of the ``fuel tank'' available to
civilization.

\begin{table}[htbp]
  \centering
  \caption{Symbols for Case~1.}
  \begin{tabular}{lll}
    \toprule
    Description                          & Symbol & Units           \\
    \midrule
    Total non-renewable energy resources & $R(0)$ & \si{\exa\joule} \\
    Non-renewable depletion time         & $t_1$  & \si{\year}      \\
    \bottomrule
  \end{tabular}
\end{table}

The derivation of the depletion time ($t_1$) begins with the boundary condition
where initial resources ($R(0)$) equal the cumulative demand ($Q$) at time
$t_1$:
\begin{equation}
  R(0) = Q(t_1)
\end{equation}
Substituting the expression for cumulative demand yields:
\begin{equation}
  R(0) = \frac{D(0)}{r}\bigl[\exp(r\,t_1) - 1\bigr]
\end{equation}
Solving for the depletion time follows the general form $t = k/r$:
\begin{align}
  t_1    &= \frac{k_1(r)}{r} \\
  k_1(r) &= \ln\!\left[\frac{R(0)}{D(0)/r} + 1\right]
\end{align}

This algebraic structure demonstrates a direct relationship between the growth
rate and the timeline for resource exhaustion. The logarithmic nature of the
numerator indicates that the magnitude of the initial resource stock ($R(0)$)
has a diminishing impact on the depletion date as the growth rate ($r$)
increases.

\subsection{Case 2: Kelvin Limit}

This section explores a thermodynamic boundary to growth based on waste heat
rather than fuel scarcity. Even with an infinite energy source, the laws of
physics dictate every unit of energy used eventually becomes waste heat. On a
planetary scale, this heat must be radiated into space to maintain a stable
environment.

Earth stays at a life-sustaining temperature by balancing received solar energy
with infrared radiation emitted back into the cosmos. Human energy consumption
from ``terrestrial'' sources, such as nuclear fusion or fossil fuels, adds
``new'' heat to this balance. Because this energy was not already part of the
solar-to-earth flow, the planet must reach a higher temperature to increase its
radiative cooling capacity and shed the additional load.

The ``Kelvin Limit'' defines the point where this added waste heat pushes
Earth's surface temperature to \SI{373}{\kelvin} (\SI{100}{\celsius}), the
boiling point of water. At this threshold, the planet becomes physically
uninhabitable.

The physics of this limit relies on the Stefan--Boltzmann Law~\cite{Pie10},
stating the power a planet radiates ($P_r$) is proportional to the fourth power
of its absolute temperature ($T$).

\begin{table}[htbp]
  \centering
  \caption{Symbols for Case~2.}
  \begin{tabular}{lll}
    \toprule
    Description                & Symbol     & Units                                     \\
    \midrule
    Radiated power             & $P_r(T)$   & \si{\exa\joule\per\year}                  \\
    Stefan--Boltzmann constant & $\sigma_B$ & \si{\watt\per\meter\squared\per\kelvin^4}  \\
    Albedo                     & $a$        & Dmnl                                      \\
    Boiling time               & $t_2$      & \si{\year}                                \\
    \bottomrule
  \end{tabular}
\end{table}

The solar power arriving at the Earth's cross-section is
$L\,\alpha \approx \SI{5474e3}{\exa\joule\per\year}$. A portion of this energy
is immediately reflected into space by the Earth's albedo ($a$), which
represents the planet's reflectivity. Based on NASA data, Earth's
albedo~\cite{NASA} is approximately 0.30, meaning 30\% of incoming light
reflects away while the remaining 70\% is absorbed as heat ($P_a$).

The model assumes an initial equilibrium where absorbed solar power ($P_a$)
equals the power radiated at the baseline temperature ($T_0$).
\begin{align}
  P_r(T)   &= \sigma\,T^4\,(4\pi\,X_e^2) \\
  P_r(T_0) &= L\,\alpha\,(1-a) = P_a
\end{align}

The limit occurs at time $t_2$, when the combined heat of the Sun ($P_a$) and
human energy demand ($D$) requires the Earth to reach the boiling point ($T_2$)
to maintain equilibrium.
\begin{equation}
  D(t_2) = P_r(T_2) - P_a
\end{equation}
Solving for $t_2$ yields a solution in the form $t = k/r$. The rate coefficient
($k_2$) quantifies the logarithmic ``thermal room'' remaining before waste heat
exceeds the radiative capacity of the planet at \SI{373}{\kelvin}.
\begin{align}
  t_2 &= \frac{k_2(r)}{r} \\
  k_2 &= \ln\!\left[\frac{L\,\alpha\,(1-a)}{D(0)}
         \left(\left[\frac{T_2}{T_0}\right]^4 - 1\right)\right]
\end{align}

This thermal wall represents a hard physical limit. Technological efficiency
cannot bypass it; higher energy use to drive AI or industry simply accelerates
the transition toward this planetary boiling point.

\subsection{Case 3: Renewable Supply on Earth Only}

This scenario examines the upper boundary of planetary energy collection from
renewable sources. Unlike non-renewable stocks, renewable energy is a ``flow''
rather than a ``tank.'' The absolute maximum power available to a terrestrial
civilization is the total solar flux intercepted by the Earth's cross-section.

Saturation occurs at time $t_3$ when instantaneous energy demand ($D$) matches
this total incoming solar power ($S$). At this point, even with 100\% efficient
collection over the entire area of the planet's shadow, the civilization can no
longer sustain exponential growth from terrestrial solar resources.

\begin{table}[htbp]
  \centering
  \caption{Symbols for Case~3.}
  \begin{tabular}{lll}
    \toprule
    Description              & Symbol & Units                    \\
    \midrule
    Solar power at Earth     & $S$    & \si{\exa\joule\per\year} \\
    Renewable depletion time & $t_3$  & \si{\year}               \\
    \bottomrule
  \end{tabular}
\end{table}

The limit is reached when the demand function intersects the solar constant
($L\alpha$):
\begin{equation}
  D(t_3) = S = L\,\alpha
\end{equation}
Substituting the exponential demand model and solving for time:
\begin{align}
  t_3 &= \frac{k_3}{r} \\
  k_3 &= \ln\!\left[\frac{L\,\alpha}{D(0)}\right]
\end{align}

\subsection{Case 4: Dyson Limit}

This scenario represents the transition from a planetary civilization to a
Kardashev Type~II civilization. At this threshold, all terrestrial energy
options, including both finite fuel stocks and the total solar flux intercepting
Earth's cross-section, have been exhausted. To sustain continued exponential
growth, civilization must expand its collection infrastructure beyond the planet
to capture the total radiant power of the Sun~\cite{NK64,Dys60}.

A Dyson sphere is a hypothetical megastructure constructed around a star to
capture its immense energy output. Originally proposed by physicist Freeman
Dyson in 1960, the most practical design is not a solid, rigid shell, which
would be mechanically impossible and unstable, but rather a ``Dyson Swarm.''
This swarm architecture would consist of a vast, distributed network of
independent solar collectors or mirrors orbiting the host star. As these
satellites harvest the star's solar radiation, they would wirelessly transmit
the collected power back to planets, space stations, or other outposts using
advanced technologies like laser beams or microwave radiation~\cite{Dys60,Bar24}.

The ``Dyson Limit'' occurs at time $t_4$, when instantaneous energy demand
($D$) matches the total solar luminosity ($L$).

\begin{table}[htbp]
  \centering
  \caption{Symbols for Case~4.}
  \begin{tabular}{lll}
    \toprule
    Description          & Symbol & Units                    \\
    \midrule
    Sun luminosity       & $L$    & \si{\exa\joule\per\year} \\
    Dyson depletion time & $t_4$  & \si{\year}               \\
    \bottomrule
  \end{tabular}
\end{table}

The limit is reached when the demand function intersects the solar luminosity:
\begin{equation}
  D(t_4) = L
\end{equation}
\begin{align}
  t_4 &= \frac{k_4}{r} \\
  k_4 &= \ln\!\left[\frac{L}{D(0)}\right]
\end{align}

\subsection{Case 5: Asimov Limit}

This final scenario represents the transition from a stellar civilization to a
Kardashev Type~III civilization~\cite{NK64}. Having harvested the total radiant
power of the Sun, further exponential growth requires expansion beyond the solar
system to capture the energy of other stars within the galaxy. The ``Asimov
Limit'' defines the point where the physical expansion of civilization's
boundary is constrained by the universal speed limit: the speed of light ($c$).

To sustain a constant energy growth rate $r$, the civilization must encompass
new stars at an accelerating pace. This expansion is modeled as a spherical
volume of space with radius $X_c$ growing over time.

\begin{table}[htbp]
  \centering
  \caption{Symbols for Case~5.}
  \begin{tabular}{lll}
    \toprule
    Description           & Symbol & Units                                          \\
    \midrule
    Civilization boundary & $X_c$  & \si{ly}                                        \\
    Local stellar density & $L/V$  & \si{\exa\joule\per\year\per ly^3}              \\
    Speed of light        & $c$    & \si{ly\per\year}                               \\
    Asimov depletion time & $t_5$  & \si{\year}                                     \\
    \bottomrule
  \end{tabular}
\end{table}

The total energy available at any time depends on the number of stars contained
within the expansion volume. If $L/V$ represents the average stellar luminosity
per unit of galactic volume~\cite{CO17}, the total power $D(t)$ is:
\begin{equation}
  D(t) = \frac{L}{V}\cdot\frac{4}{3}\pi X_c^3
\end{equation}

To maintain exponential growth, the radius $X_c$ must increase. Differentiating
this demand shows that the required velocity of the expansion front
($dX_c/dt$) is proportional to the growth rate and the current radius. Because
no physical influence or resource collection craft can travel faster than light,
the expansion velocity has an absolute ceiling:
\begin{equation}
  \frac{dX_c}{dt} \leq c
\end{equation}

The Asimov Limit occurs at time $t_5$, when the growth rate $r$ requires an
expansion velocity equal to the speed of light. At this juncture, the
civilization can no longer add new stars fast enough to satisfy the compounding
demand of its existing population~\cite{Asi56}.

Substituting the maximum possible expansion velocity into the demand growth
equations yields the solution for $t_5$ in the standard $t = k/r$ form. The
rate coefficient ($k_5$) incorporates the light-speed constraint and the
density of stars in the galactic neighborhood:
\begin{align}
  t_5    &= \frac{k_5(r)}{r} \\
  k_5(r) &= \ln\!\left[\frac{L}{D(0)}\frac{4\pi/3}{V}
             \left(\frac{c}{r/3}\right)^3\right]
\end{align}

Because $k_5$ includes the growth rate $r$ within the logarithmic term, higher
growth rates will eventually hit the ``light speed wall''~\cite{Wri20}.

\subsection{Numerical Example}

When we map our projected AI energy demand against these physical ceilings, the
results are startling. By using a baseline demand ($D(0)$) of
\SI{633}{\exa\joule\per\year} and a fuel tank ($R(0)$) built on global
deuterium reserves, we can calculate the ``logarithmic distance'' to each
limit, represented by the rate coefficients ($k$) in Table~\ref{tab:kcoef}.
While some of these coefficients shift slightly with the growth rate ($r$), they
essentially serve as a set of fixed hurdles humanity must clear as it expands.

\begin{table}[htbp]
  \centering
  \caption{Calculated Rate Coefficients.}
  \label{tab:kcoef}
  \begin{tabular}{lrrrrr}
    \toprule
    & $k_1$ & $k_2$ & $k_3$ & $k_4$ & $k_5$ \\
    $r\;(\si{\per\year})$ & Non-Renew. & Kelvin & Renew. & Dyson & Stellar \\
    \midrule
    0.010 & 19.3 & 10.0 & 9.1 & 30.6 & 43.6 \\
    0.015 & 19.7 & 10.0 & 9.1 & 30.6 & 42.4 \\
    0.030 & 20.4 & 10.0 & 9.1 & 30.6 & 40.3 \\
    0.10  & 21.6 & 10.0 & 9.1 & 30.6 & 36.7 \\
    0.15  & 22.0 & 10.0 & 9.1 & 30.6 & 35.5 \\
    \bottomrule
  \end{tabular}
\end{table}

The real impact of AI-driven growth becomes clear when we look at the actual
timescales for these transitions. Table~\ref{tab:timelines} shows the depletion
times ($t = k/r$) across various growth scenarios. At the historical industrial
growth rate of $\SI{0.01}{\per\year}$, we see the slow-motion expansion Kardashev originally
envisioned: it takes \SI{3000}{\year} to reach the Dyson Sphere stage. However,
when we apply growth rates typical of the AI revolution ($\SI{0.15}{\per\year}$), those millennia
collapse into decades.

\begin{table}[htbp]
  \centering
  \caption{Energy Depletion Timelines.}
  \label{tab:timelines}
  \begin{tabular}{lrrrrr}
    \toprule
    $r\;(\si{\per\year})$ & $k=10$ & $k=20$ & $k=30$ & $k=40$ & $k=50$ \\
    $t\;(\si{\year})$ & & & & & \\
    \midrule
    0.010 & 1000 & 2000 & 3000 & 4000 & 5000 \\
    0.015 &  667 & 1333 & 2000 & 2667 & 3333 \\
    0.030 &  333 &  667 & 1000 & 1333 & 1667 \\
    0.10  &  100 &  200 &  300 &  400 &  500 \\
    0.15  &   67 &  133 &  200 &  267 &  333 \\
    \bottomrule
  \end{tabular}
\end{table}

These figures represent a fundamental shift in the prospects of civilization
under AI. Under the 15\% growth rates currently demonstrated by AI
infrastructure, we quickly accelerate past all projected limits. The thermal
``Kelvin Limit'' ($k \approx 10$), which would normally take ten centuries to
reach, suddenly appears in just \SI{67}{\year}, well within a single human lifetime.

Even the most optimistic expansion scenarios fail to buy much time. Moving from
the surface of the Earth to harvesting the total power of the Sun ($k \approx
30$) adds only \SI{133}{\year} of headroom. Even the leap to a galactic-scale
Type~III civilization ($k \approx 50$) provides only a few centuries of growth
before expansion is halted by the speed of light.

%%=============================================================================
\section{Simulation}
%%=============================================================================

\begin{figure}[htbp]
  \centering
  \includegraphics[width=0.75\columnwidth]{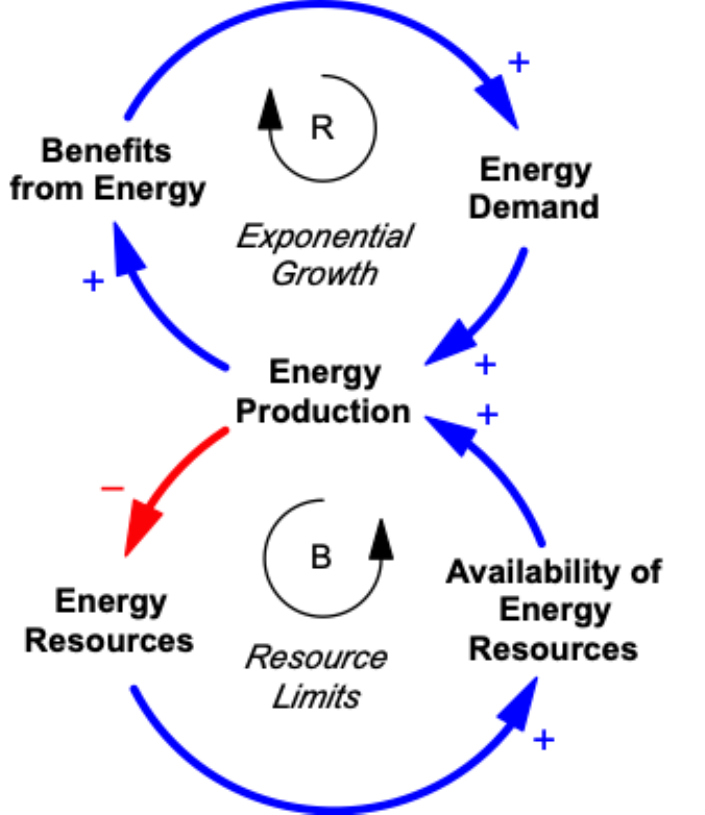}
  \caption{Causal Loop Diagram of the Basic Feedback Structure of the Energy
           Model.}
  \label{fig:CLD}
\end{figure}

A System Dynamics model is presented to capture the transition of terrestrial
energy resources. Figure~\ref{fig:CLD} illustrates the causal loop diagram of
the basic terrestrial energy feedback structure with two competing loops. Energy
demand growth stems primarily from the benefits of energy production creating a
reinforcing feedback loop driving higher demand. However, this growth eventually
meets the physical constraints of energy production that depends on the
availability of energy resources, which are consumed during the process. When
resources become scarce, energy production slows, reducing the benefits and
dampening further demand.

\begin{figure}[htbp]
  \centering
  \includegraphics[width=\columnwidth]{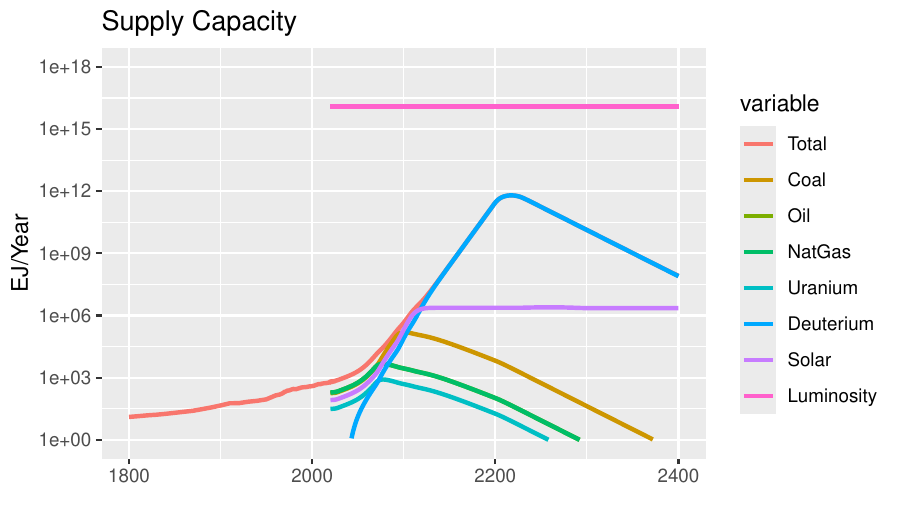}
  \caption{Terrestrial Energy Supply.}
  \label{fig:supply}
\end{figure}

Real-world energy supply utilizes a mix of renewable and non-renewable sources.
When a non-renewable resource reaches depletion, supply naturally shifts to
alternative resources to meet demand. Figure~\ref{fig:supply} presents the
results of the energy model simulation.

The model assumes energy demand grows at $\SI{0.02}{\per\year}$ for non-AI
scenarios and $\SI{0.15}{\per\year}$ for AI-driven scenarios. As coal, oil,
natural gas, and uranium are consumed, solar and fusion capacity expand to meet
demand. Solar panels and deuterium fusion grow in tandem until approximately
2100, when solar panels cover the terrestrial surface. In this scenario, supply
tracks demand growth until the planet deuterium reserves are depleted in 2200.
Slowing the energy demand growth rate shifts these curves further into the
future, yet the fundamental shapes and outcomes remain the same.

If AI-driven growth in energy demand continues, then building solar collection
in space is the obvious next step.

\begin{figure}[htbp]
  \centering
  \includegraphics[width=\columnwidth]{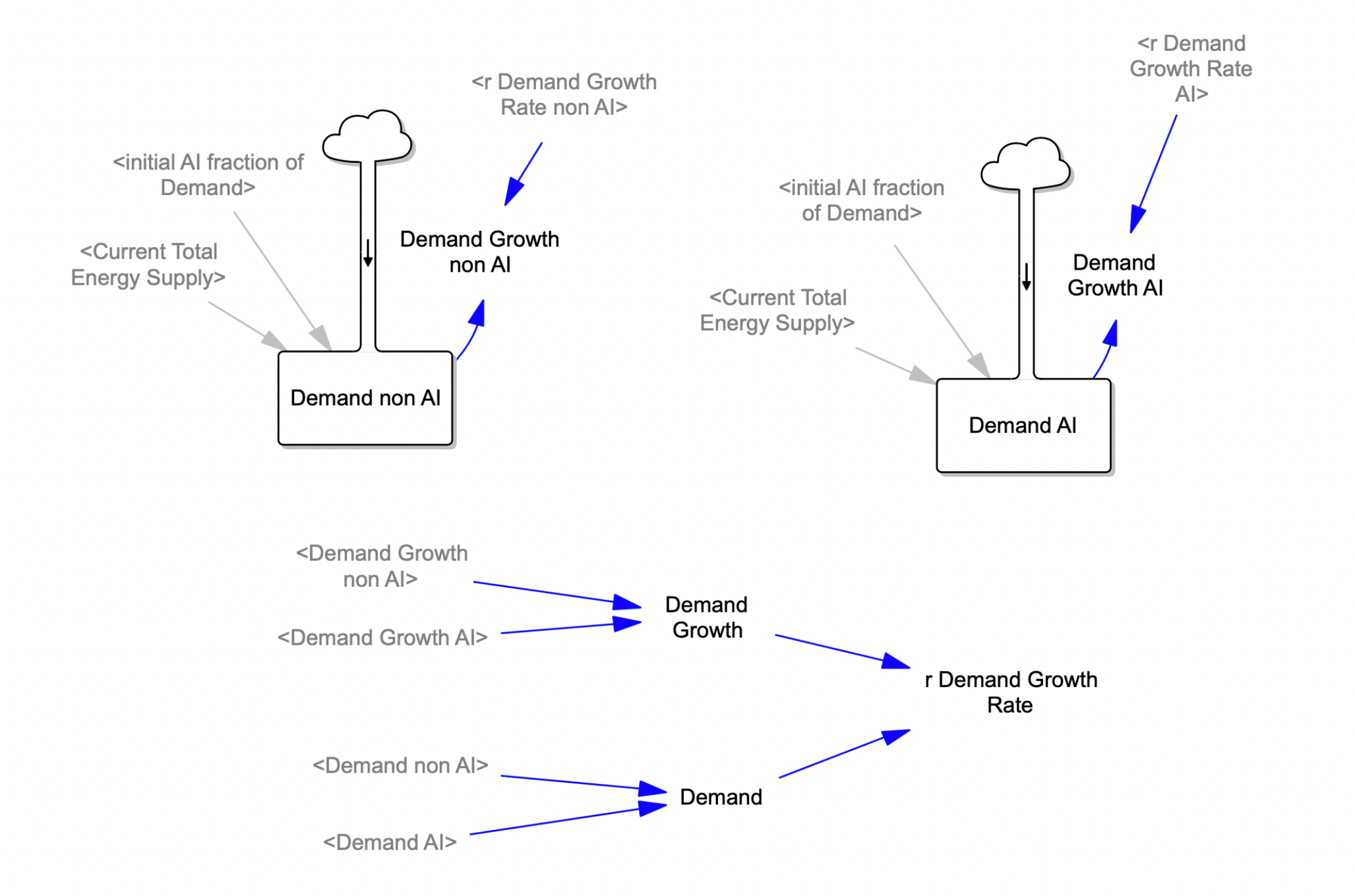}
  \caption{Stocks and Flows of Energy Demand.}
  \label{fig:demand_sd}
\end{figure}

Figure~\ref{fig:demand_sd} illustrates the modeling of energy demand. In this
simulation, growth rates for AI and non-AI energy remain exogenous constants.
Even with constant growth rates, the rate of flow into the respective demand
stocks increases alongside the level of the demand stock. Higher demand
inherently leads to faster expansion.

\begin{figure}[htbp]
  \centering
  \includegraphics[width=\columnwidth]{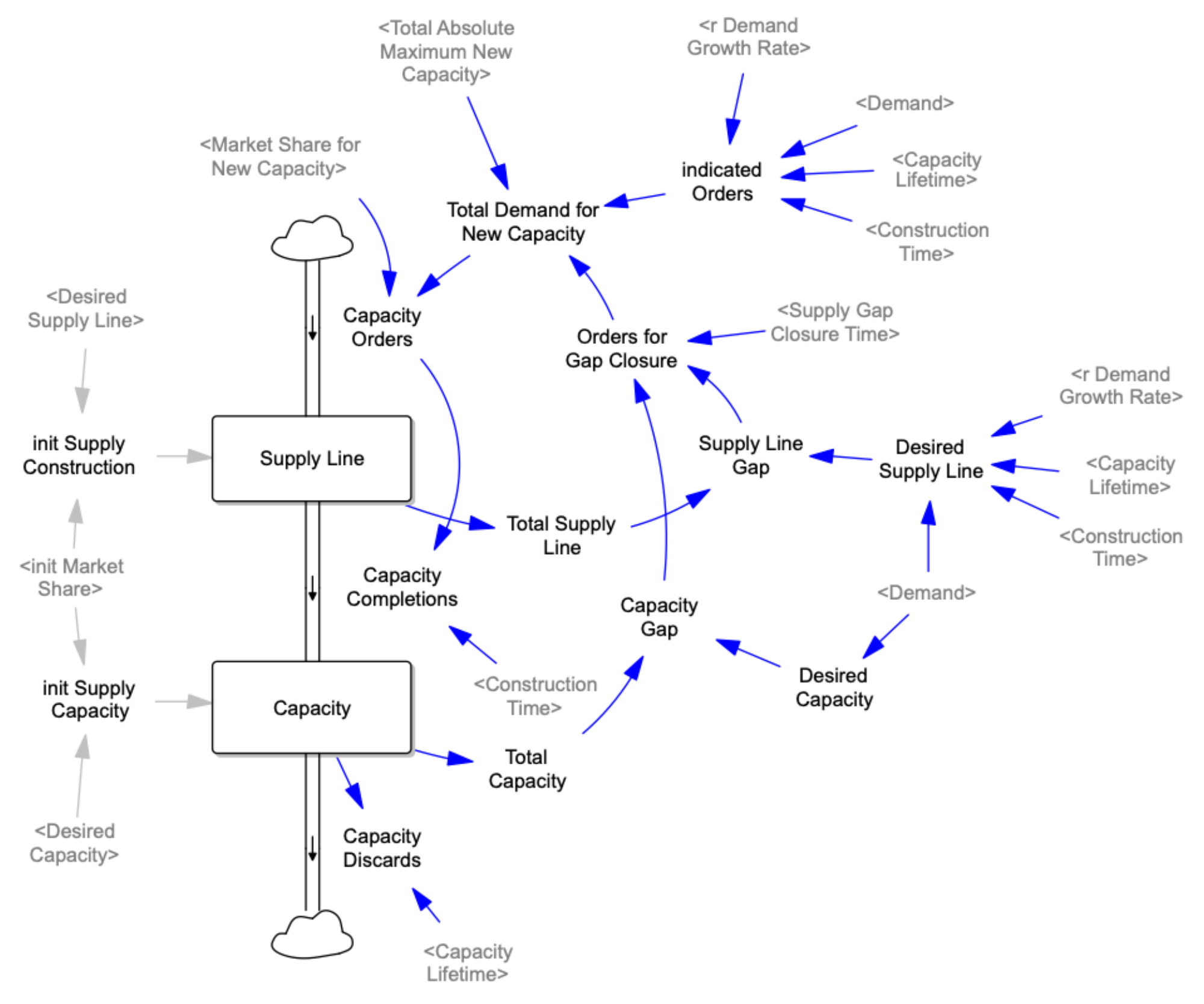}
  \caption{Capacity for Energy Production.}
  \label{fig:capacity}
\end{figure}

Figure~\ref{fig:capacity} presents the model for new energy production.
Increasing demand triggers orders for oil pumps, coal mines, nuclear plants, and
solar panels. These orders fill the Supply Line stock. Once constructed, these
facilities increase the total Capacity stock of energy production. Eventually,
these facilities undergo decommissioning due to resource exhaustion or age,
which reduces the overall capacity.

\begin{figure}[htbp]
  \centering
  \includegraphics[width=\columnwidth]{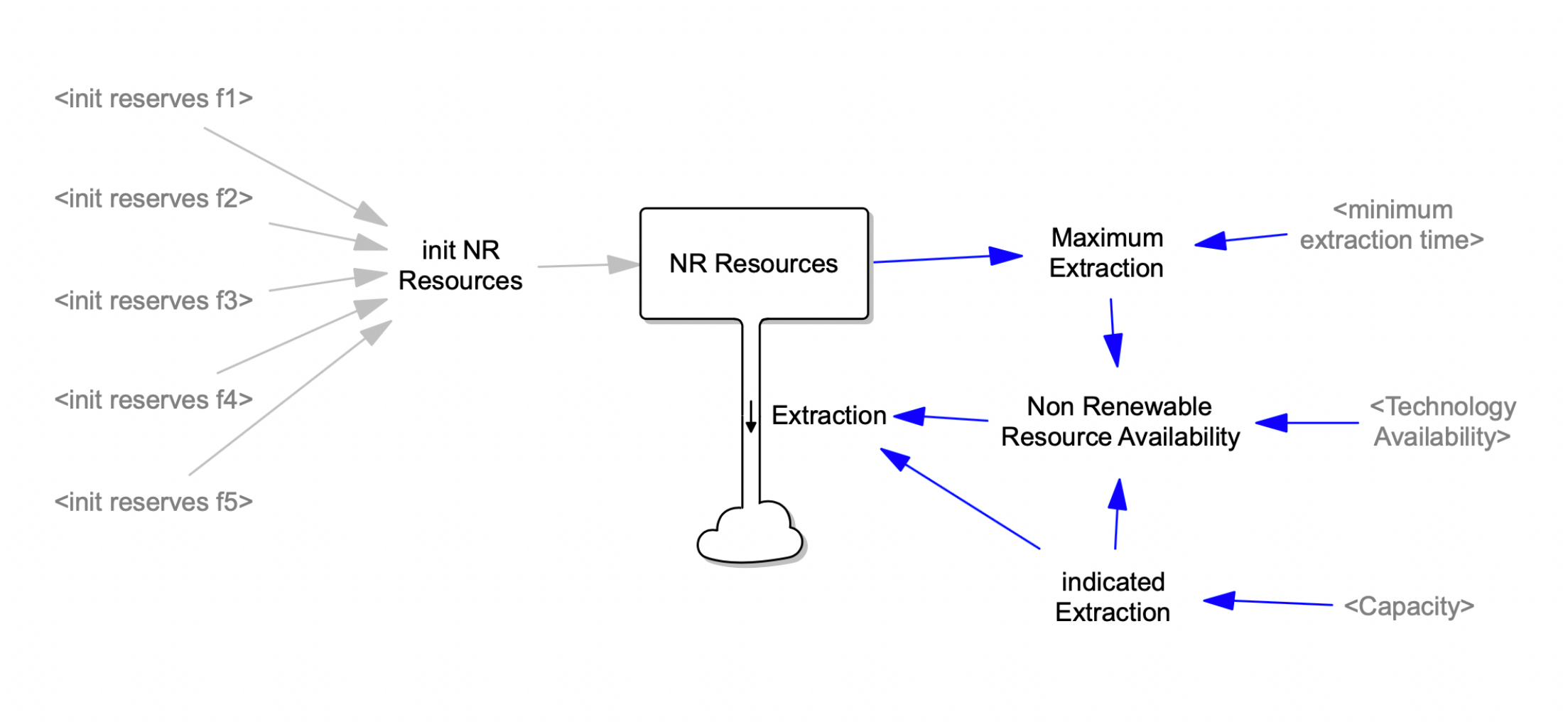}
  \caption{Non-Renewable Resources.}
  \label{fig:nonrenew}
\end{figure}

The final major component of the model, shown in Figure~\ref{fig:nonrenew},
represents the depletion of non-renewable resources. Terrestrial stocks of oil,
coal, natural gas, uranium, and deuterium are finite. Each extraction point
represented in the capacity model drains these respective resource stocks.
Eventually, the non-renewable energy stocks reach exhaustion, forcing the energy
supply to shift entirely to renewable sources.

%%=============================================================================
\section{Discussion}
%%=============================================================================

The consequences of constant exponential growth are profoundly
counter-intuitive. While traditional forecasts estimate depletion by comparing
current resources to current demand, the mathematical models in this paper
demonstrate how growth radically compresses these timelines. This acceleration
renders historical precedents nearly irrelevant when projecting the future of
AI-driven expansion.

Non-renewable resources face exhaustion within a remarkably short period. This
remains true even if current estimates of fossil fuel reserves are off by
several orders of magnitude. Regardless of the shift toward renewable energy,
the sheer magnitude of compounding demand creates an overwhelming pressure to
consume every available terrestrial fuel source. This trajectory carries severe
implications for global climate stability, as the speed of consumption likely
outpaces the planetary capacity to sequester carbon.

The commercialization of fusion represents a significant variable in these
projections. However, the exponential surge in AI power requirements makes the
model largely insensitive to the precise timing of fusion availability. While
different fusion fuel cycles offer varying efficiencies, the conclusion remains
unchanged: civilization eventually exhausts even the vast deuterium reserves of
the Earth's oceans.

Several internal mechanisms may naturally decelerate the growth of AI before it
hits these physical walls. These balancing loops include:

\begin{itemize}

\item \textit{Economic Constraints:} Beyond initial capital expenditures, energy
prices represent the primary operating cost for AI. If supply gaps drive
electricity costs higher, the diminishing profitability of AI services will
naturally slow infrastructure expansion.

\item \textit{Efficiency Gains:} AI is already optimizing the design and
manufacturing of the next generation of hardware. While total demand increases,
improvements in computational efficiency and non-AI operations may provide
temporary offsets. These optimizations become a physical necessity for any
infrastructure intended for deep-space exploration.

\item \textit{Architectural Shifts:} The industry is moving from power-hungry
training phases to more efficient inference-heavy models~\cite{Cla26}. While
training requires massive bursts of energy, individual inference tasks demand
far less power because the model simply applies existing knowledge to new
inputs~\cite{Cloud}. This shift prioritizes cost-efficiency and real-time
responsiveness over raw computational force~\cite{Aro25}.

\item \textit{The Data Wall:} AI developers face the impending exhaustion of
high-quality, human-generated public text data~\cite{Bev26}. Historically, AI
progressed by feeding exponentially more data into larger models, but experts
estimate the stock of effective training data could be consumed by
2032~\cite{Vil24}. Without new breakthroughs in synthetic data or autonomous
knowledge generation, the ``bigger is better'' approach faces a ceiling of
diminishing returns.

\end{itemize}

Driven by these staggering cooling and energy demands, tech giants and aerospace
startups are already drafting plans to move data centers into
orbit~\cite{Don26,Rev26}. Companies like SpaceX and LoneStar Data Holdings
envision utilizing near-constant solar flux and the natural vacuum of space for
free radiative cooling~\cite{Barg,Del25}. This migration bypasses terrestrial
constraints such as strained power grids and massive water consumption.
Space-based computing also establishes the essential digital backbone for lunar
settlements and future deep-space missions~\cite{Del25}.

The limits described in this paper may sound like science fiction, yet
contemporary engineering is rapidly closing the gap. Scientific breakeven in
fusion has been demonstrated~\cite{DOE}, and commercial deployment is actively
pursued~\cite{CFS}. Blue Origin~\cite{BO} and SpaceX~\cite{SX} have developed
the largest rockets in history to facilitate the colonization of the Moon and
Mars. SpaceX recently applied for regulatory approval to launch a constellation
of AI data centers~\cite{FCC}, explicitly citing the transition toward a
Kardashev Type~II civilization as a primary motivation.

In the last century, humanity discovered quantum mechanics, general relativity,
and nuclear energy. While these pillars of physics remain unreconciled and the
nature of dark matter remains largely a mystery~\cite{NASA2}, new discoveries
may yet rewrite the boundaries of the possible. However, as General Gordon
Sullivan observed, ``hope is not a strategy''~\cite{Sul97}. Given the extremely
short timescales identified in our models, civilization must plan its expansion
using the rigorous laws of physics known today.

%%=============================================================================
\section{Conclusion}
%%=============================================================================

The physical quantities examined in this analysis, including the total deuterium
reserves within the oceans, the full luminosity of the Sun, and civilization
expanding at the speed of light, initially seem unthinkably vast. Yet, because
these values function as variables within a logarithmic framework, they become
remarkably finite when confronted by sustained exponential growth. This research
demonstrates any positive growth rate in energy demand inevitably exhausts
available solar system resources within a relatively compressed timeframe. The
remaining window for terrestrial or even stellar expansion is directly
proportional to the reciprocal of the growth rate; consequently, slower growth
provides humanity more time to navigate these transitions.

The central challenge involves the fundamental nature of exponential expansion
rather than the specific numerical value of the growth rate itself. While the
rapid adoption of Artificial Intelligence significantly hastens these limits,
the underlying physical hurdles remain fixed by planetary geometry and universal
laws. As energy requirements begin to decouple from traditional economic trends,
the transition toward a Kardashev Type~II or Type~III civilization shifts from
a distant theoretical exercise to a looming engineering necessity. To ensure
long-term stability, global energy strategies must account for these absolute
ceilings. Civilization must eventually reconcile its desire for perpetual growth
with the rigid thermodynamic and relativistic boundaries of the cosmos.

%%=============================================================================
%% References
%%=============================================================================
%% References ordered by first appearance in text.

%%=============================================================================
\appendix
%%=============================================================================

\section{Deuterium Resources}
\label{sect:deuterium}

Since non-renewable resources are dominated by deuterium, and estimates of deuterium resources are not as widely available as fossil fuels and uranium, we document the details here.  There are two parts to the estimate: (1) inventory of deuterium in moles, and (2) fusion energy per mole.

We start with the concentration of deuterium in seawater~\cite{IAEAd}.
\begin{align}
\SI{33}{\gram\per\cubic\meter}
&= \frac{(\SI{33}{\gram\per\cubic\meter})}{(\SI{2.014}{\gram\per\mol})}
\left( \frac{\SI{1e3}{\meter}}{\SI{1}{\kilo\meter}} \right)^3 \nonumber \\
&= \SI{16.4e9}{\mol\per(\kilo\meter)^3}
\end{align}
We multiply by the volume of seawater~\cite{NOAA} to obtain the inventory of deuterium:
\begin{align}
&\left(\SI{16.4e9}{\mol\per(\kilo\meter)^3}\right)
 \left(\SI{1386e6}{(\kilo\meter)^3}\right) \nonumber \\
&\quad = \SI{22.7e18}{\mol}
\end{align}

The full deuterium cycle is given by NRL~\cite{NRL}
\begin{align}
\ch{^2D + ^3T &-> ^4He + ^0n + 17.6 MeV} \\
\ch{^2D + ^2D &-> ^3T + ^1p + 4.03 MeV} \\
\ch{^2D + ^2D &-> ^3He + ^1n + 3.27 MeV} \\
\ch{^2D + ^3He &-> ^4He + ^1p + 18.3 MeV} \\
\ch{5 ^2D &-> ^4He + 2 ^0n + ^3He + ^1p + 24.9 MeV} \\
\ch{6 ^2D &-> 2 ^4He + 2 ^0n + 2 ^1p + 43.2 MeV}
\end{align}
Equation~(32) applies if we neglect the D-$^3$He reaction; Eq.~(33) applies if we include it.

If we include the D-$^3$He reaction, the energy release per deuteron is $(43.2/6)$~\si{MeV}.
\begin{align}
\SI{1}{\mol.D}
&= (\SI{1}{\mol.D})
\left( \frac{\num{6.02e23}}{\si{\mol}} \right)
\left(\frac{\SI{1.602e-13}{\joule}}{\si{MeV}} \right) \nonumber \\
&\quad\times \left( \frac{43.2}{6}~\si{MeV} \right) \nonumber \\
&= \SI{694}{\giga\joule\per\mol}
\end{align}

Putting it all together, we estimate the resources of deuterium by
\begin{align}
R(0) &\approx (\SI{22.7e18}{\mol}) \, (\SI{694}{\giga\joule\per\mol}) \nonumber \\
     &= \SI{15.8e12}{\exa\joule}
\end{align}

\end{document}